\documentclass[prb,twocolumn,float,showpacs,preprintnumbers,amsmath,amssymb,superscriptaddress]{revtex4}
\usepackage{graphicx}
\usepackage{dcolumn}
\usepackage{bm}
\begin{document}
\title{ Paramagnetic effect in
YBa$_2$Cu$_3$O$_{7-x}$ grain boundary junctions.}
\author{E.Il'ichev}
\email{ilichev@ipht-jena.de}
\affiliation{%
Institute for Physical High Technology, P.O. Box 100239, D-07702
Jena, Germany}
\author{F. Tafuri}
\affiliation{Gruppo collegato INFM-Dip. Ingegneria
dell'Informazione, Seconda Universit\`{a} di Napoli, 81031 Aversa
(CE), Italy} \affiliation{IBM Watson Research Center, Route 134
Yorktown Heights, 10598 NY, USA}
\author{M. Grajcar}%
\altaffiliation[On leave from ]{Department of Solid State Physics,
Comenius University, SK-84248 Bratislava, Slovakia.}
\affiliation{Friedrich Schiller University, Institute of Solid
State Physics, D-07743 Jena, Germany.}
\author {R.P.J. IJsselsteijn}
\affiliation{%
Institute for Physical High Technology, P.O. Box 100239, D-07702
Jena, Germany}
\author {J. Weber}
\affiliation{%
Institute for Physical High Technology, P.O. Box 100239, D-07702
Jena, Germany}

\author{F. Lombardi}
\affiliation{INFM - Dip. Scienze Fisiche, Universit\`{a} di Napoli
Federico II, 80125 Napoli, Italy}
\author{J.R. Kirtley}
\affiliation{IBM Watson Research Center, Route 134 Yorktown
Heights, 10598 NY, USA}
\date{\today}
\input epsf
\begin{abstract} { A detailed investigation of the magnetic response of
YBa$_{2}$Cu$_{3}$O$_{x}$ grain boundary Josephson junctions has
been carried out using both radio-frequency measurements
and Scanning SQUID Microscopy. In a nominally zero-field-cooled
regime we observed a paramagnetic response at low external fields
for 45$^\circ$ $asymmetric$ grain boundaries. We argue that the
observed phenomenology results from the $d$-wave order parameter
symmetry and depends on Andreev bound states.}
\end{abstract}
 \maketitle
Superconductors below the transition temperature $T_{c}$ usually
expel an external magnetic field. This phenomenon, known as the
Meissner effect, leads to diamagnetism. It was therefore quite
unexpected that  a paramagnetic Meissner effect (PME) was observed
in a field cooled regime for ceramic
Bi$_{2}$Sr$_{2}$CaCu$_{2}$O$_{8}$ \cite {Braunish,Heinzel}
(BiSCCO). For an explanation, it was proposed that below $T_{c}$
there are randomly distributed spontaneous orbital currents in the
sample.\cite {Braunish,Kus} These currents can be oriented by an
external magnetic field, providing a paramagnetic signal. Sigrist
and Rice \cite {Sig} pointed out that the spontaneous orbital
currents can be caused by a $d_{x^{2}-y^{2}}$-wave symmetry of the
superconducting state. Indeed, the $d$-wave scenario predicts the
existence of Josephson junctions in which the Josephson energy
reaches a minimum for a phase difference across the junction at
$\phi=\pi$ ($\pi$-contact). In ceramic high-$T_c$ superconductors
(HTS) the grains can form loops which contain odd numbers of
$\pi$-contacts, so called frustrated loops.\cite{Sig} In such a
system the energy is minimized by a configuration with a
spontaneous current flowing in the loop.\cite{changjohn} In
particular, for a $\pi$-contact with a ``conventional"
current-phase relationship $I=I_{c} \sin (\phi+\pi)$ inserted in a
loop, the gain in Josephson energy exceeds the loss of magnetic
energy when the parameter $\beta$ satisfies $\beta =2\pi L
I_c/\Phi_0 > 1$, where $\Phi_{0}$ is the flux quantum, $I_{c}$ is
the critical current and $L$ is a loop inductance. However, a
paramagnetic signal in the field-cooled regime has  also been
observed for conventional superconductors.\cite{Thomp,Geim}

In order to distinguish between different origins of paramagnetism
Rice and Sigrist \cite{Rice} proposed to detect spontaneous
orbital currents by a SQUID microscope for zero-field cooled
samples. In a granular BiSCCO sample exhibiting a paramagnetic
signal, spontaneous magnetization has indeed been
observed.\cite{Kir}
The PME and the presence of spontaneous currents therefore
represent two of the main features induced by the $d$-wave order
parameter symmetry in HTS.

A study of these phenomena on a more controlled system, such as a
grain boundary (GB) line, could shed additional light on the
relation between these effects and their influence on the
properties of the grain boundary Josephson junctions (GBJJ).


In this paper, for the first time, we give evidence  of a
paramagnetic behavior for zero-field cooled asymmetric 45$^\circ$
YBa$_{2}$Cu$_{3}$O$_{x}$ (YBCO) GBJJ.
The paramagnetic signal, that has been previously observed in
two-dimensional systems, has been in this case detected in a
single GB line and related to basic mechanisms in HTS JJs.  Apart
from conventional transport properties and Scanning SQUID
Microscopy (SSM) measurements, we exploited radio-frequency
measurements, which are very sensitive and provide a direct test
for paramagnetic signals as demonstrated below. This is a further
manifestation of the $d$-wave nature of the order parameter, but
it is also relevant for the understanding of transport across
GBJJs and in particular on the incidence of Andreev bound states.

The idea of the $rf$ measurements is the following. The sample of
interest is inductively coupled to a high-quality (quality factor
$Q\approx 300$) parallel resonant circuit of inductance $L_T$,
capacitance $C_T$. Experimentally this is realized by flip chip
configuration - the sample is placed on the top of a small solenoid
coil perpendicular to its axis. For such an arrangement, the
effective impedance of the tank circuit coupled to the sample is a
function of the external magnetic field $H_e$ applied to the sample.
This field consists of a $dc$ and an $rf$ component as induced by a
current $I_{dc}$ (in fact of very low frequency) and an $rf$
oscillating current $I_{rf}$ in the tank coil, hence $H_e = H_{dc} +
H_{rf}$. In order to avoid the nucleation of $rf$ ``vortices" and
other nonlinear effects, the amplitude of $H_{rf}$ is small, so that
$H_{rf}\ll H_{c1}$,$H_{dc}$ where $H_{c1}$ is the first critical
field, therefore $H_e \cong H_{dc}$. If $L_{eff}(H_{dc})$ and
$R_{eff}(H_{dc})$ are the effective inductance  and the effective
resistance of the tank circuit-sample system respectively, the phase
angle $\alpha$ between the drive current $I_{rf}$ and the tank
voltage $U$ is given by
\begin{equation}
\tan\alpha=\frac{1}{R_{eff}(H_{dc})}\cdot\left(\frac{1}{\omega
C_T}-\omega L_{eff}(H_{dc})\right). \label{Eq:alpha}
\end{equation}
It follows from Eq.~\ref{Eq:alpha} for $H_{dc}=0$ and at the
resonance frequency $\omega_0 = 1/\sqrt{L_{eff}(0)C_T}$ that the
parameter $\alpha$ is zero.  Therefore, by monitoring $\alpha$ as
a function of $H_{dc}$ at the frequency $\omega_0$, the
$L_{eff}(H_{dc})$ dependence can be obtained (note that
$R_{eff}(H_{dc})$ dependence is controlled independently, by
measuring the quality factor $Q$ of the tank circuit-sample
system).  If $R_{eff}= R_{T}$ does not depend on $H_{dc}$ then tan
$\alpha$ can be written as:
\begin{equation}
\tan\alpha=-k^2Q\chi_m \label{Eq:M}
\end{equation}
where $k$ is the coupling coefficient between the tank coil and
the sample, and $\chi_m$ is ac magnetic susceptibility of the
sample. In the experiments we are obviously interested in the
regime $\chi_m\neq const$, the only configuration able to provide
significant information. Let us analyze Eq.~\ref{Eq:M} near $H_e =
0$. When the supercurrent induced by an externally applied field
is diamagnetic, a change of $\chi_m<0$ will result in a local
maximum in $\alpha (H_e)$. Similarly, a local minimum of the
$\alpha (H_e)$ curve indicates a paramagnetic response
($\chi_m>0$).

The simplest system which exhibits a similar behavior is the
well-known $rf$ SQUID.  The sensor of this device is a Josephson
junction inserted in a superconducting loop. If the inductance $L$
of the loop is relatively small (so that $\beta<1$), the $\alpha
(H_{dc})$ dependence has a local maximum at $H_{dc}=0$ (see for
example Ref.~\onlinecite{Mal}). Due to the induced current in the
sensor, the magnetic flux $\Phi_{i}$ inside the loop satisfies
$\Phi_{i}<\Phi_{e}$, where $\Phi_{e}$ is an applied external flux.
Therefore, the magnetic response is diamagnetic. If, instead of a
conventional junction, a $\pi$-contact is inserted in the same
sensor, the $\alpha (H_{dc})$ dependence has a local minimum at
$H_{dc}=0$ and $\Phi_{i}>\Phi_{e}$, providing a paramagnetic
response. The phase shift $\alpha$ can be obtained from
Eq.~\ref{Eq:M} defining
$\chi_m=d\Phi_i/d\Phi_e-1$.\cite{Ilichev00} The value of $k^{2}Q$
allows the estimation of the resolution of the $rf$ measurement,
in this case of the order of a few percent of the flux quantum.

As we discussed above, within the framework of the $d$-wave
scenario, extrinsic effects such as faceting can play a relevant
role in the determination of the properties of the junctions and
can contribute in particular to causing spontaneous currents
and/or a paramagnetic effect. In the case of a strongly meandering
interface, the GB exhibits a random parallel array of $0$- and
$\pi$-contacts. This is somehow the analog of the explanation of
the PME in BiSCCO crystals in the 1-dimensional case of a GB line
($\pi$-loops model). An alternative mechanism of the PME can be
given in terms of the midgap  states (MGS) and surface properties
of {\it d}-wave superconductors (MGS model). In this case, the MGS
model is valid even for flat interfaces of a {\it d}-wave
superconductor.

The interplay between the PME and superconductivity, apart from being
an interesting topic itself on $d$-wave induced effects, may be
crucial to improve understanding of transport properties of GBs. It
has been demonstrated that GB transport properties strongly depend on
the quality of the substrate,\cite{Jia} of the thin film and the type
of growth. This led to apparently conflicting behaviors in a wide
spectrum of transport regimes. Zig-zag Nb-Au-YBCO junctions isolated
the effect of facets by investigating the presence of spontaneous
currents and anomalous magnetic patterns.\cite{Hans1} We intend to
reach the other limit by investigating a morphology with nominal very
reduced faceting, and low barrier transparency. To this aim we
employed biepitaxial junctions, where ``clean" basal plane GBs can be
reproducibly obtained.

\begin{figure}[t]
\vspace{9pt}
\centering\includegraphics*[width=0.60\linewidth,angle=-90]{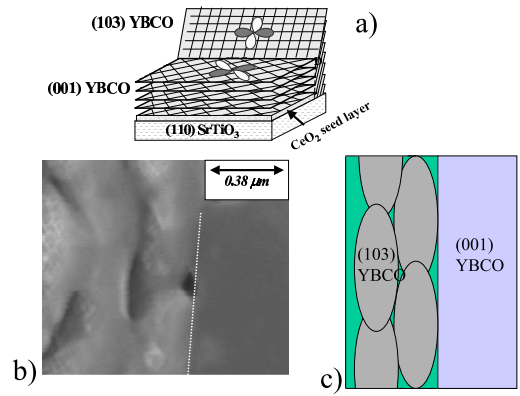}
\caption{\label{fig:SEM} a) Sketch of the investigated biepitaxial
grain boundary where a 45$^\circ$ $c$-axis tilt accompanies the
45$^\circ$ $a$-axis tilt; in b) a Scanning Electron Microscope
image relative to the this GB is shown. The elongated grains on
the left typical of the (103) growth may generate clean interfaces
with reduced faceting, as evident also by the sketch in c).}
\end{figure}

The biepitaxial technique allows the fabrication of various GBJJs
by growing different seed layers and using substrates with
different orientations.\cite{tafuri,tafurib} In this experiment we
have used CeO$_2$ as a seed layer material deposited on (110)
SrTiO$_3$ substrates. Details of the fabrication procedure can be
found elsewhere.\cite{tafurib} YBCO grows along the [001]
direction on CeO$_2$ seed layers, while it grows along the
[103]/[013] direction on SrTiO$_3$ substrates. By using CeO$_2$ as
a seed layer we were able to induce a 45$^\circ$ rotation of the
$a-b$ plane of the YBCO with respect to the in-plane direction of
the SrTiO$_3$ substrate, and as a consequence the $\pi$ contact is
formed.\cite{tafuriprl} The measurements shown below are for the
case where a 45$^\circ$ $c$-axis tilt accompanies the 45$^\circ$
$a$-axis tilt (Fig.~\ref{fig:SEM}a.) This configuration in
principle leads to interfaces where effects due to faceting can be
very reduced, as shown for instance in the Scanning Electron
Microscope image of Fig.~\ref{fig:SEM}b, and the relative sketch
Fig.~\ref{fig:SEM}c. For biepitaxial junctions in the tilt
case,\cite{tafuri} the YBCO growth kinetics and the junction
interface orientation determine that the long side of the [103]
grains faces the $c$-axis counter-electrode, and this leads to a
more controlled GB (basal plane GB). This has been confirmed by
cross section Transmission Electron Microscope
investigations.\cite{tafuri}

The sample had originally a typical $SQUID$ geometry with a central hole of 50 $\mu$m $\times$
50 $\mu$m. Rf and $SSM$ measurements were performed on a configuration where one of
junctions was removed and the other was reduced to about 60 $\mu$m.

\begin{figure}[t]
\vspace{9pt}
\centering\includegraphics*[width=0.90\linewidth]{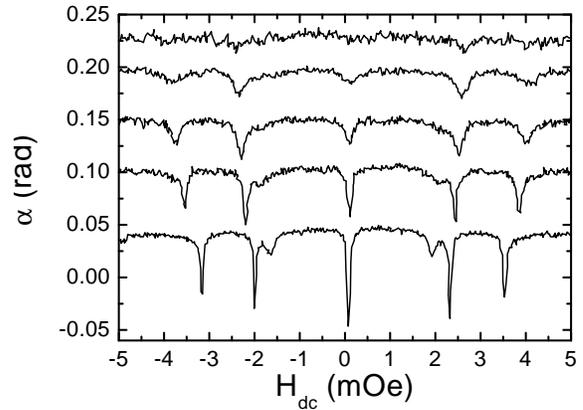}
\caption{\label{fig:alpha}The phase angle $\alpha$ as a function
of the external magnetic field  applied parallel to $c$-axes of
65$\mu$m-wide bridge across an asymmetric 45$^{0}$
YBa$_{2}$Cu$_{3}$O$_{x}$ grain boundary. From top to bottom the
data correspond to $T$=40, 30, 20, 10, and 4.2 K. The data are
vertically shifted for clarity.}
\end{figure}

Results of $rf$ measurements as a function of an externally applied
magnetic field at the temperature range from $T$= 4.2 K up to 40~K
are presented in Fig.~\ref{fig:alpha} for nominally zero field cooled
samples (the rest field is below 0.2 mOe). The sharp minimum at
$H_{dc}$=0 in Fig.~\ref{fig:alpha} represents an unusual feature with
respect to most GB systems, which exhibit a maximum for zero field.
The minima at a finite magnetic field are originated by the
redistribution of the magnetic flux into the GB as extensively
discussed in Ref.~\onlinecite{Sige}.

\begin{figure}[t]
\vspace{9pt}
\centering\includegraphics*[width=0.90\linewidth,angle=0]{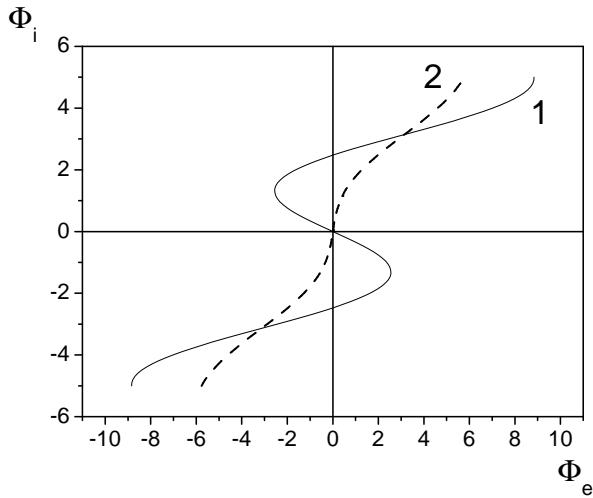}
\caption{Schematic drawings - time averaged magnetic flux inside
the GB vs applied magnetic flux (1) - without thermal fluctuation,
(2) - large thermal fluctuation.} \label{fig:fig4}
\end{figure}

According to the discussion above, the minimum of $\alpha$ at
$H_{dc}=0$ is direct evidence of paramagnetic behavior of this
type of GB. Furthermore the paramagnetic response was absent after
the removal of YBCO from the GB region, demonstrating that the
effect is only due to the GB. In other words we just repeated the
same measurement on the same sample after removing only a narrow
region of YBCO along the grain boundary line in order  to prove
that the paramagnetic signal was caused by the grain boundary and
not by any uncontrolled environmental reason (substrate or sample
holder) or possible impurities in the YBCO thin film far from the
GB line. The absence of hysteresis for $\alpha
(H_{dc})$-dependence with respect to external field means that
there is no spontaneous surface current or flux generated in the
GB. This is analogous to the situation in $rf$ SQUID with
$\pi$-contact for  $\beta <1$. As a matter of fact, in the case of
finite fluctuations the jumps of $\Phi_i$ can occur with a certain
probability when $\Phi_e$ falls in a hysteresis region.\cite{kur}
When the distribution width of the jump probability is larger than
the width of the hysteresis, the flux jumps many times during the
measurements. As a result the apparent time-averaged $\Phi_i
(\Phi_e)$ dependence presents a finite slope rather than
hysteresis even for $\beta >1$ (see Fig.~\ref{fig:fig4}). This
situation has already been observed experimentally.\cite{il98}

A detailed insight into the problem of the paramagnetic effect and of
spontaneous currents in facetted GB's has been given in
Refs.~\onlinecite{Hans,Mints}. In this model the current density
$j_{S}(x)$ ($x$ is the coordinate in the plane of the contact) is a
random and alternating function of $x$, $j_{c}(x)=\langle
j_{c}\rangle[1+g_{0}f(x)]$ with $\langle f(x)\rangle =0$, max[
$f(x)$]=1 and $g_{0}=$ max[$j_{c}(x)/\langle j_{c}\rangle$]. The
length scale $l$ of $g(x)$ variations is of the order of the grain
boundary meandering, typically $l$ is in the range of 0.01-0.1
$\mu$m.\cite{Hans} For a conventional current-phase relationship
$j_{c}(x)=\langle j_{c}\rangle \sin\varphi (x)$, the properties of
the GBJJ are determined by the  parameter\cite {Mints}
$\gamma=g_{0}^{2}l^{2}/8\pi^{2}\Lambda_{j}^{2}$. Here
$\Lambda_{j}=(c\Phi_{0}/16\pi^{2}\lambda\langle j_{c}\rangle )^{1/2}$
is the effective Josephson penetration depth and $\lambda$ is the
London penetration depth. Within this model, for $\gamma<1$ there is
no created flux in GBJJ and the microjunctions stay in a ``excited
currentless state".\cite {Cop} In principle, such junctions can
exhibit a paramagnetic response. When $\gamma>1$ there is flux
generated in the GBJJ. Qualitatively, this requirement is similar to
the condition $\beta>1$ for the $rf$ SQUID, as discussed above.  Thus
the same arguments concerning thermal fluctuations can be done for
our experiments. If at $\Phi_e =0$ and $T = 0$ there is spontaneous
flux in the GB, one should observe hysteresis of the $\Phi_i
(\Phi_e)$ dependence (see Fig.~\ref{fig:fig4}, curve 1). However, at
finite temperature fluctuations wash out the hysteresis, therefore
there is no spontaneous current (see Fig.~\ref{fig:fig4}, curve 2).
The above scenario relies on a macroscopic approach to the GB,
which is a random array of parallel 0 and $\pi$-junctions, $i.e.$
the ``extrinsic" effect of faceting.

The other possible scenario is based on microscopic arguments, and
mainly on mechanisms based on Andreev reflection at the
surface.\cite{wendin, Tan, hasci, low1} The $d$-wave symmetry of
the gap leads to the formation of the surface state at zero energy
(so-called midgap states - MGS \cite{Hu94}). This MGS is
degenerate with respect to the direction along the surface of the
superconductor ($\pm k_{y}$). Any mechanism which is able to split
the MGS will lower the energy of the system.\cite{sig} The
splitting of the MGS leads to spontaneous currents along the
surface and therefore to time reversal symmetry breaking.

Different mechanisms of the MGS splitting were proposed in the
literature (see Ref.~\onlinecite{low1} and Refs. therein). Apart
from suggestions based on the presence of an imaginary component
of the order parameter (like $d \pm is$, see for example
Ref.~\onlinecite{am}), there are at least two scenarios based only
on a pure d-wave symmetry of order parameter.

The first scenario takes into account the Josephson effect. For
asymmetric 45$^\circ$ GBJJ the energy of the MGS can be written
as\cite{wendin}
\begin{equation}
\varepsilon_{\rm MGS}(\varphi,k_y) = -{\rm sgn}
(k_y)E_0(\vartheta)\sin (\varphi) , \label{Eq:MGS}
\end{equation}
where
$E_0=\Delta_L|\Delta_R|D(\vartheta)/(2|\Delta_L|+D(\vartheta)
[|\Delta_R|-|\Delta_L|])$, $D(\vartheta)$ is the angle-dependent
barrier transparency and $\Delta_L$, $\Delta_R$ are
superconducting energy gaps for the angle $\vartheta=\arcsin
k_y/k_F$ in the left and right superconductor, respectively. For
asymmetric 45$^\circ$ GBJJ the Josephson energy minimum
corresponds to the equilibrium phase difference across the
junction $\varphi =\pi/2$ which leads to the splitting of the
MGS.\cite{Ilichev01} Hence, the spontaneous paramagnetic current
flows in the surface layer $\sim \xi_0$. This current is screened
by the Meissner supercurrent on the scale of $\lambda$ and the
system pays for the gain in the Josephson energy at the expense of
the energy of the magnetic field.

Within the framework of the second scenario the free surface of a
$d$-wave superconductor has been found to be responsible for the
appearance of the paramagnetic effect. In this case the paramagnetic
quasi-particle current tries to compensate the Meissner
supercurrent.\cite{hasci} As a matter of fact the supercurrent causes
a redistribution of quasi-particles in ${\bf k}$-space due to the
shift $E_{\bf k}=E_{\bf k}^0+{\bf p}_{\bf k}\cdot {\bf v}_s$ in the
quasiparticle excitation energy, where $E^0_{\bf k}$ is the
excitation energy in the absence of a supercurrent, ${\bf v}_s$ is
the velocity  of the supercurrent and ${\bf p}_{\bf k}$ is momentum
of the quasiparticle. This is obviously relevant for a (110) oriented
HTS surface due to the existence of the MGS.

The key feature for both of the described approaches is the
presence of a mechanism able to split the MGS and produce a gap
$E_0$, or in other words to populate the $\pm k_y$ MGS unequally.
This phenomenon is accompanied by a phase transition to a broken
time reversal symmetry (TRSB) state. The transition temperature
$T_{TRSB}$ can be evaluated and in both  approaches $T_{TRSB}$
have been estimated to be below 1~K, in agreement with tunneling
experiments.\cite{Covington97} $T_{TRSB}$ will be apparently the
relevant parameter also for the PME. However, in analogy with $rf$
SQUID properties,  above $T_{TRSB}$ one should observe no
hysteresis but steep slope of $\Phi_i (\Phi_e)$ at $\Phi_e=0$
(again see Fig.~\ref{fig:fig4}). This explains why we observed PME
in a wide temperature range well above the $T_{TRSB}$ values but
no hysteretic behavior.

\begin{figure}[t]
\vspace{10pt}
\centering\includegraphics*[width=0.50\linewidth,angle=-90]{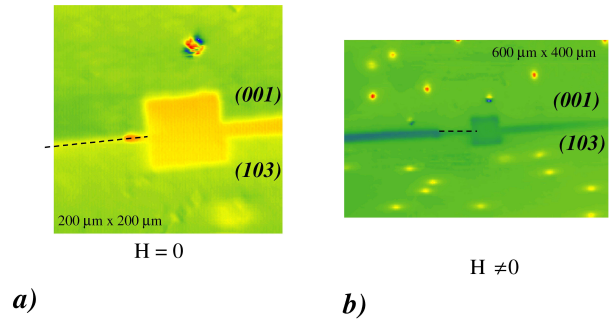}
\caption{\label{fig:SSM} Scanning SQUID Microscopy image of a)a 200
$\times$ 200 $\mu$m$^2$ area, enclosing the tilt GB; the sample was
cooled in zero magnetic field; b) a 600 $\times$ 400 $\mu$m$^2$ area,
enclosing the tilt GB; the sample was cooled in a small magnetic
field. Both images refer to the sample measured by rf techniques, and
the grain boundary junction is indicated by a dotted line. In all
images, taken at $T$ = 4.2 K, there is no evidence of spontaneous
currents. In b) the superconducting nature of both electrodes is
proven by the presence of Abrikosov vortices, which appear elongated
in the (103) electrode.}
\end{figure}

Additional information has been provided by SSM investigations, which
revealed the absence of spontaneous currents, as shown in the SSM
images of the GBJJ region in Fig.~\ref{fig:SSM}.
Fig.~\ref{fig:SSM}(a) and Fig. ~\ref{fig:SSM}(b) are SSM images of a
wide area around the GB "in zero field cooling, and in non-zero field
cooling respectively, both at $T$=4.2 K. Figure~\ref{fig:SSM}(a)
confirms in zero field cooling the absence of any spontaneous
magnetization for this sample. This result is different from other
measurements on 45$^\circ$ asymmetric bicrystal \cite{Man} and
biepitaxial GBJJ.\cite{kirtley1} We attribute this difference to the
concomitance of reduced faceting along the GB and the low barrier
transparency of this junction, characterized by low values of the
critical current density $j_C$ about
$10^{2}$A/cm$^{2}$.\cite{tafuriprb} Fig.~\ref{fig:SSM} (b) gives
evidence of vortices in both the electrodes, and in particular of
anisotropic vortices in the (103) electrode. As a consequence, it
seems to be not the case that the PME is originated by $\pi$-loops
across the GB ($\pi$-loops model, see above). Moreover, the observed
narrow dip on $\alpha\ vs.\ H_{dc}$ characteristics at $H_{dc}=0$
(see Fig.~\ref{fig:alpha}) requires $\gamma\approx 1$. Simple
estimations show that $\gamma\approx 1$ corresponds to critical
current densities across the GBJJ of the order of $10^{5}$A/cm$^{2}$,
which seems to be unrealistic, given that the average critical
current density for our junction are of the order of
$10^{2}$A/cm$^{2}$. On the other hand the MGS model is apparently
consistent with our results.

In conclusion, we have measured the magnetic-field response of an
asymmetric $45^{0}$ grain boundary in a YBa$_{2}$Cu$_{3}$O$_{x}$
thin film. The results of these investigations allow us to
identify Andreev bound states as the cause of the paramagnetic
effect, confirming theoretical predictions. Andreev bound states
have been studied in detail theoretically in $HTS$ Josephson
junctions and systems. However, due to their extreme localization
and stringent survival conditions, experimental detection has been
basically confirmed only by one type of measurement, zero bias
anomalies in tunneling spectra.\cite{Covington97} Our method is
complementary and direct, and relies on the comparison of various
types of measurements realized on the same samples. The
contribution to the existing state of knowledge on d-wave order
parameter symmetry in $HTS$ superconductors is to demonstrate
experimentally the occurrence of bound states through an
innovative approach, shedding light on the paramagnetic effect in
a novel topological configuration.

A paramagnetic response at low external fields was found in a
zero-field-cooling regime. We have shown that the observed
phenomena can be explained by the paramagnetic response of MGS
surface currents.

The authors would like to thank V. Zakosarenko, A. Golubov and Yu.
Barash for numerous illuminating discussions.  E.I. and M.G. were
partially supported by the D-wave Systems Inc. M.G. wants to
acknowledge the support of Grants No. VEGA 1/9177/02 and
APVT-51-021602.
This work has been partially supported by the ESF project "$\Pi$-Shift".

\end{document}